\documentclass[final]{eps}

\usepackage{graphicx}
\usepackage{times}
\usepackage{amsmath,amssymb}

\volumenumber{00}
\publishyear{2012}
\frompage{\pageref{firstpage}}
\topage{\pageref{finalpage}}

\shorttitle{R. S. Asano \lowercase{\textit{et al.}}: Dust formation
history of galaxies}

\title{Dust formation history of galaxies: a critical role of
metallicity\footnotemark for the dust mass growth by accreting materials in the interstellar
medium}
\author{Ryosuke S. Asano$^1$, Tsutomu T. Takeuchi$^1$, Hiroyuki
Hirashita$^2$ and Akio K. Inoue$^3$}
\affiliation{$^1$Department of Particle and Astrophysical Science, Nagoya University, Furo-cho, Chikusa-ku, Nagoya 464-8602, JAPAN\\
             $^2$Academia Sinica Institute for Astronomy and Astrophysics, P.\ O.\ Box 23-141, Taipei 10617, TAIWAN\\
             $^3$College of General Education, Osaka Sangyo University, 3-1-1 Nakagaito, Daito, Osaka 574-8530, JAPAN}
\received{xxxx xx, 2011}\revised{xxxx xx, 2012}\accepted{xxxx xx, 2012}
\abstract{This paper investigate what is the main driver of the dust mass growth
   in the interstellar medium (ISM) by using a chemical evolution model
   of galaxy with metals (elements heavier than helium) in dust phase in addition
   to the total amount of metals.
   We consider asymptotic giant branch (AGB) stars, type II supernovae (SNe~II)
   and the dust mass growth in the ISM as the sources of dust, and SN
   shocks as the destruction mechanism of dust.
   Further, to describe the dust evolution precisely, 
   our model takes into account the age and metallicity (the ratio of metal mass to ISM
   mass) dependence of the sources of dust.
   We particularly focused on the dust mass growth, and found that the dust mass growth in the ISM
   is regulated by the metallicity.
   To quantify this aspect, we introduce a ``critical metallicity'', which is a metallicity 
   at which the contribution of stars (AGB stars and SNe~II) equals that of the dust mass growth in
   the ISM.
   If the star formation timescale is shorter, the value of the critical
   metallicity is higher, but the galactic age at which the metallicity
   reaches the critical metallicity is shorter. 
   From observations, it was expected that the dust mass growth was the dominant source
   of dust in the Milky Way and dusty QSOs at high redshifts.
   By introducing the critical metallicity, it is clearly
   shown that the dust mass growth is the main source of dust in
   such galaxies with various star formation timescales and ages.
   The dust mass growth in the ISM is regulated by metallicity, and
   we stress that the critical metallicity works as an indicator to judge whether the grain
   growth in the ISM is dominant source of dust in a galaxy,
   especially because of the strong and nonlinear dependence on the 
   metallicity.}
\keywords{dust, extinction --- galaxies: infrared --- galaxies: evolution
--- galaxies: starburst --- stars: formation}

\begin{document}
\label{firstpage}
\maketitle
\copyrighttext{}

\section{Introduction}%\label{sec:intro}

Stellar light, in particular at shorter wavelengths, is absorbed by dust
and re-emitted as the far-infrared thermal emission from the dust 
(e.g., Witt \& Gordon, 2000, and references therein). 
Therefore dust affects the spectral energy distributions of galaxies 
(e.g., Takagi et al., 1999; Granato et al., 2000; Noll et al., 2009; Popescu et al., 2011).
The existence of dust in galaxies also affects the star formation
activity. 
Dust grains increase the molecular formation
rate by two orders of magnitude compared to the case without dust
(e.g., Hollenbach \& McKee, 1979), 
and the interstellar medium (ISM) is cooled efficiently by molecules and dust. 
Consequently, star formation is activated drastically by dust.
Hence, dust is one of the most important factors for the evolution of galaxies
(e.g., Hirashita \& Ferrara, 2002; Yamasawa et al., 2011). 
\footnotetext{${}^*$the ratio of metal (elements heavier than helium) mass to ISM mass}

The amount of dust in galaxies is one of the crucial factors to interpret
the observational information of galaxies, since dust exists ubiquitously and 
the radiation from stars is always affected by dust attenuation. 
However, in spite of its importance, the evolution of dust
amount has not been completely established yet.
There are some key factors to understand dust evolution of galaxies.
One of the keys is the ratio of metal (elements heavier than
helium) mass to ISM mass, which is called ``metallicity''.
Since dust grains consist of metals, it is natural to think
that the evolution of dust is closely related to metallicity.
In general, galaxies are believed to evolve from the state with a very
low metallicity and very small amount of dust to higher amounts of metal
and dust.  
Hence, it is a mandatory to model the formation and evolution of dust grains
in galaxies along with the evolution of metallicity 
(e.g., Dwek \& Scalo, 1980; Hirashita, 1999a,b; Inoue, 2003; Yamasawa et
al., 2011). 

Dust grains are formed by the condensation of metals.
A significant part of the metals released by stellar mass loss during stellar evolution
or supernovae (SNe) at the end of the life of stars condense into dust grains.
Dust grains are not only supplied from stars but also destroyed by SNe
blast waves (e.g., Jones et al., 1994; Jones, Tielens \& Hollenbach,
1996; Nozawa et al., 2003; Zhukovska, Gail \& Trieloff, 2008). 
In addition, we should consider the dust mass growth in the ISM by the
accretion of atoms and molecules of refractory elements onto grains
(e.g., Dwek, 1998; Liffman \& Clayton, 1989; Draine, 2009; Jones \&
Nuth, 2011).

What kind of dust formation processes are dominant at each stage of galaxy evolution?
It is a very important question for understanding the evolution history of the ISM
and star formation in galaxies.
However, since dust evolution depends strongly on the age 
and metallicity of a galaxy,
it is not easy to answer this question.
Up to now, dust evolution has been studied with various models.
For example, in young galaxies, SNe 
have been considered as the source of dust because they are the final stage
of massive stars whose lifetime is short, and asymptotic giant branch
(AGB) stars have been neglected because of their longer lifetime. 
However, Valiante et al. (2009) showed that the AGB stars also contribute to
the dust production in young galaxies and cannot be neglected even
on a short timescale of $\sim 500$~Myr.
A more elaborate survey of the parameter space for the dust formation by
SNe and AGB stars has been done by Gall et al. (2011a).
They showed that the contribution of AGB stars exceeds that of
SNe~II at several $100$~Myr if the ratio between metal and dust mass
produced by SNe~II is less $\sim$ 0.01 and mass-heavy IMF with
mass range $1$--$100\;\mbox{M}_{\odot}$.

As for the dust mass growth, the ISM is considered to be the
main source of dust in various galaxies.
For example, the present dust amount observed in the Milky Way cannot be explained
if the source of dust would have been only stars, suggesting that we
must consider the dust mass growth in the ISM in evolved
galaxies (e.g., Dwek, 1998; Liffman \& Clayton, 1989; Draine, 2009; Jones \&
Nuth, 2011).  
Recently, dusty quasars (total dust mass $> 10^8\;{\rm
M}_{\odot}$) have been discovered at high redshifts
(e.g., Beelen et al., 2006; Wang et al., 2008), 
and theoretical studies on dust sources at high redshifts are currently 
carried out actively
(e.g., Micha\l owski et al., 2010b; Gall et al., 2011a,b; Pipino et al.,
2011; Valiante et al., 2011).
They showed that it is hard to explain the total dust
amount in these QSOs only with stellar contributions, and then they
discussed the importance of the dust mass growth in the ISM.
Next question is what controls the point where the dust mass growth in the
ISM becomes efficient to the total dust mass in galaxies.
Therefore, although each physical process has been already extensively
discussed in preceding studies, there emerges a crucial question: 
what kind of dust production process is dominant at each stage of galaxy evolution?
Particularly, when the dust dust mass growth becomes dominant as a source of
dust mass?

The central aim of this work is to address this question.
In this paper, we investigate what is
the main driver of the dust mass growth in the ISM.
Since all sources of dust production are tightly related to each 
other on dust evolution, it is crucial to treat these processes in a
unified framework to understand the evolution of dust in galaxies.
Here, we adopt the model based on chemical evolution model in a same manner
as Hirashita (1999b); Calura, Pipino \& Matteucci (2008); Inoue (2011).
This is because their models consider main dust
production/destruction processes that affect the dust evolution of
galaxies, and it is easy to compare our results to the previous ones.
{}From this work, we find that the dust mass growth in the ISM is
regulated by metallicity.
We refer to this metallicity as the critical
metallicity, whose details are described in Sect.~3.2.
Although the dust mass growth can occur at any time of a galaxy age, 
but we stress this point because there is a moment at which the
dust mass growth overwhelms the contribution from other sources 
of dust. 

This paper is organized as follows.
In Sect.~2, we describe the model developed for this work.
In Sect.~3,  
we show and discuss the basic results obtained by our model.
The main topic of this paper, critical metallicity, is introduced and
extensively examined in Sect.~3.2.
Section~4 is devoted to the conclusions.
The solar metallicity is set to be ${\rm Z}_{\odot} =
0.02$ (Anders \& Grevesse, 1989) throughout this paper.

\section{Dust evolution model of galaxies}
%\label{sec:model}

In this section, we show the simple chemical evolution model with dust
which examine what determines the point where the dust mass growth in the ISM
becomes effective.
The dust evolution model is built with the same manner as in Hirashita (1999b) and
Inoue (2011).

\subsection{Equations of galaxy evolution}
%\label{subsec:equation}

In this subsection, we describe the equations of mass evolution of stars
and the ISM which contains metal and dust in galaxies.
We treat a galaxy as one-zone
because we are interested in global properties of galaxies.
Also, we assume a closed-box model. 
Thus, the total baryon mass $M_{\rm
tot}$ (the sum of the stellar mass and the ISM mass) is a constant. 
However, since $M_{\rm tot}$ is just a scale factor in our model, this
value does not affect the physical properties of galaxies nonlinearly.

In this work, we do not consider the effects of inflow and outflow.
However, they may not influence the properties of dust and metal
enrichment in galaxies because of the following reasons: 
An inflow makes not only metallicity but also dust-to-gas mass
ratio small, because usually an inflow is considered to be metal- and dust-poor. 
As for an outflow, it expels ISM components (gas, metal and
dust) out of a galaxy.
However, if all ISM components flow out together,
the metallicity and dust-to-gas mass ratio do not change. 

Under these settings, equations of the time evolution of the total
stellar mass $M_*$, ISM mass $M_{\rm ISM}$, metal mass $M_Z$, and dust
mass $M_{\rm d}$ are (e.g., Lisenfeld \& Ferrara, 1998; Hirashita, 1999b) 
\begin{eqnarray}
\frac{{\rm d}M_*(t)}{{\rm d}t} &=& \mbox{SFR}(t) - R(t),\\
%\label{eq:ismevo}
\frac{{\rm d}M_{\rm ISM}(t)}{{\rm d}t} &=& -\mbox{SFR}(t) + R(t),\\
%\label{eq:timemetal}
\frac{{\rm d}M_Z(t)}{{\rm d}t} &=& -Z(t)\mbox{SFR}(t) + R_Z(t) +
 Y_Z(t),\\
\nonumber \frac{{\rm d}M_{\rm d}(t)}{{\rm d}t} &=& - {\cal D}(t)\mbox{SFR}(t) +
 Y_{\rm d}(t) - \frac{M_{\rm d}}{\tau_{\rm SN}} +
 \eta\frac{M_{\rm d}(1 - \delta)}{\tau_{\rm acc}},\\
%\label{eq:timeacc}
\end{eqnarray}
where SFR is the star formation rate, $Z(t) \equiv M_{\rm Z}/M_{\rm ISM}$ is
the metallicity, ${\cal D} \equiv M_{\rm d}/M_{\rm ISM}$ is the
dust-to-gas mass ratio, $\delta \equiv M_{\rm d}/M_Z$ is the fraction of
the metals in dust.
$\eta$ is the mass fraction of cold clouds
where the accretion process occurs, $\tau_{\rm SN}$ and $\tau_{\rm acc}$
are the timescales of dust destruction and accretion, respectively.
The definitions of these timescales are described later.
Also, $R(t)$ and $R_Z(t)$ are the total
baryon mass returned by stars and the total metal mass once injected in stars
and just returned in the ISM per unit time, respectively.
$Y_{Z}(t)$ and $Y_{\rm d}(t)$ are the
total metal mass newly produced and ejected by stars and the total dust
mass ejected by stars per unit time, respectively.
There is another notation which $Y_Z$ includes $R_Z$ (e.g., Inoue, 2011).
In this case, the value of $Y_Z$ is different from that in this paper 
because of the different definition.

For the SFR, we adopt the Schmidt law (Schmidt, 1959); SFR $\propto M^n_{\rm ISM}$.
Here, we adopt $n = 1$ for simplicity. Thus, the SFR is expressed as
\begin{equation}
\mbox{SFR}(t) = \frac{M_{\rm ISM}(t)}{\tau_{\rm SF}},
%\label{eq:sfr}
\end{equation} 
where $\tau_{\rm SF}$ is the star formation timescale.

Also, $R(t)$, $R_Z(t)$, $Y_{Z}(t)$ and $Y_{\rm d}(t)$ are written by
\begin{eqnarray}
%\label{eq:returngas}
\nonumber R(t) &=& \int^{100\;{\rm M}_{\odot}}_{m_{\rm
 cut}(t)}[m - \omega(m,Z(t - \tau_m))]\\
&&\hspace{50pt}\times \phi(m)\mbox{SFR}(t - \tau_m){\rm d}m,\\
%\label{eq:returnmetal}
\nonumber R_Z(t) &=& \int^{100\;{\rm M}_{\odot}}_{m_{\rm
 cut}(t)}[m - \omega(m,Z(t - \tau_m))]\\
\nonumber&&\hspace{50pt}\times \phi(m)\mbox{SFR}(t - \tau_m)Z(t -
 \tau_m){\rm d}m,\\
\\
\nonumber Y_Z(t) &=& \int^{100\;{\rm M}_{\odot}}_{m_{\rm
 cut}(t)}m_Z(m,Z(t - \tau_m))\\
&&\hspace{50pt}\times \phi(m)\mbox{SFR}(t - \tau_m){\rm d}m,\\
%\label{eq:returndust}
\nonumber Y_{\rm d}(t) &=& \int^{100\;{\rm M}_{\odot}}_{m_{\rm
 cut}(t)}m_{\rm d}(m,Z(t - \tau_m))\\
&&\hspace{50pt}\times \phi(m)\mbox{SFR}(t - \tau_m){\rm d}m,
%\label{eq:stellardust}
\end{eqnarray}
where $\phi(m)$ is the initial mass function (IMF), $\omega(m,Z(t)), m_Z(m,Z(t))$ and $m_{\rm d}(m,Z(t))$ are
the remnant mass, metal mass newly produced and ejected, and dust mass
produced and ejected by a star of initial mass $m$ and metallicity $Z$, respectively. 
The lifetime of a star of initial mass $m$ is expressed as
$\tau_m$, is taken from Raiteri, Villata \& Navarro (1996).
In this work, since its metallicity dependence is weak, we always adopt
the lifetime of the case of the solar metallicity.
The lower limit $m_{\rm cut}(t)$ is the mass of a star with the lifetime
$\tau_m = t$.
As for IMF, we adopt Larson IMF (Larson, 1998) in the stellar mass range
$0.1$--$100\;{\rm M}_{\odot}$, 
\begin{equation}
\phi(m) \propto m^{-(\alpha + 1)}\exp{\left(-\frac{m_{\rm ch}}{m}\right)}.
\end{equation}
Here we adopt $\alpha = 1.35$ and $m_{\rm ch} = 0.35\;{\rm M}_{\odot}$.
Also, we normalize it as
\begin{equation}
\int^{100\;{\rm M}_{\odot}}_{0.1\;{\rm M}_{\odot}}m\phi(m){\rm d}m = 1.
\end{equation}
To calculate above four equations
[Eq.~(6)--(9)],
we quoted the data of remnant mass ($\omega(m,Z)$), metal mass
($m_Z(m,Z)$) and dust mass ($m_{\rm d}(m,Z)$) of stars with mass $m$ and
metallicity $Z$ from some of previous works.

In this work, we consider AGB stars and SNe~II as stellar sources, but
neglected the SNe~Ia for simplicity.
Nozawa et al. (2011) recently proved that SNe~Ia produce little
amount of dust.
Further, Calura, Pipino \& Matteucci (2008) showed that the dust destruction rate by SNe~Ia is about
1/10 of that by SNe~II.
As for the metals ejected by SNe~Ia, they play an important
role in the chemical evolution of galaxies (e.g., Matteucci et al., 2009).
However, since we discuss not the abundance ratio of each metal but
the total metallicity, we did not take into account the contribution of SNe~Ia.

In this paper, we assume that the mass ranges of AGB stars and SNe~II
are $1$--$8\;\mbox{M}_{\odot}$ and $8$--$40\;\mbox{M}_{\odot}$,
respectively.
Also, we assume that all stars with initial masses $m >
40\;\mbox{M}_{\odot}$ evolve to black holes without SN explosions (Heger
et al., 2003).

As for the remnant and metal masses, the data are taken from
van den Hoek \& Groenewegen (1997) for AGB stars with mass range $1$--$7\;{\rm M}_{\odot}$ and
metallicities $Z = (5.0 \times 10^{-2}, 0.2, 0.4, 1.0)\;\mbox{Z}_{\odot}$
and from Woosley \& Weaver (1995) for SNe~II with mass range $12$--$40\;{\rm
M}_{\odot}$ and metallicities $Z = (5.0 \times 10^{-2}, 0.1, 1.0)\;\mbox{Z}_{\odot}$.
As for the dust mass, the data is taken from Zhukovska, Gail \& Trieloff
(2008) for AGB
stars with mass range $1$--$7\;{\rm M}_{\odot}$ and
metallicities $Z = (5.0 \times 10^{-2}, 0.1, 0.2, 0.4, 0.75, 1.0)\;\mbox{Z}_{\odot}$
and from Valiante et al. (2009) for SNe~II with mass range $12$--$40\;{\rm
M}_{\odot}$ and metallicities $Z = (5.0 \times 10^{-2},
1.0)\;\mbox{Z}_{\odot}$ which are quoted from Bianchi \& Schneider (2007). 

Although stardust yields are not completely understood, theoretical
predictions of SNe~II recently show a good agreement with observations
of nearby supernova remnants (SNRs) (e.g., Nozawa et al., 2010). 
We considered the current model based on these latest results.
Yet some problems still remain unsolved (e.g., nucleation efficiency). 
As for the dust yield of AGB stars, we adopt similar stardust yields of
Valiante et al. (2009, 2011) and Gall et al. (2011a) while their yields may be uncertain. 
However, we stress that after the dust mass growth in the ISM becomes
effective, the dust abundance is insensitive to the stardust yields
(Inoue, 2011).
Thus, although dust yields have slight uncertainties,
 we can discuss the activation mechanism for the dust mass growth in the ISM
 without ambiguity.

\subsection{Dust destruction timescale}

It is thought that SNe are the main source of the dust destruction.
This dust destruction process depends on various parameters (density and temperature of
the ISM, the explosion energy of SNe, and etc.), is very complex
(e.g., Jones et al., 1994; Jones, Tielens \& Hollenbach, 1996; Nozawa et
al., 2006). 
In this work, we adopt the formula presented by Mckee (1989).

The timescale of dust destruction $\tau_{\rm SN}$ is expressed as
\begin{equation}
\tau_{\rm SN} = \frac{M_{\rm ISM}(t)}{\epsilon m_{\rm swept} \gamma_{\rm
 SN}(t)},
\end{equation}
where $\epsilon$ is the efficiency of dust destruction by SN shocks, and
is defined as the ratio of the destroyed dust to the total swept dust by
SN shocks, $m_{\rm swept}$ is the swept ISM mass by a SN shock,
$\gamma_{\rm SN}(t)$ is the SN rate.
In this work, we assume $\epsilon = 0.1$ (Mckee, 1989; Nozawa et al., 2006).

The SN rate $\gamma(t)$ is expressed as
\begin{equation}
\gamma_{\rm SN}(t) = \int^{40\;{\rm M}_{\odot}}_{m_{\rm cut}(t) >
 8\;{\rm M}_{\odot}} \phi(m)\mbox{SFR}(t - \tau_m){\rm d}m.
\end{equation}
The range of the integration is the mass range where the SNe can occur
(Heger et al., 2003).
So, if $t < \tau_{40\;{\rm M}_{\odot}}$, $\gamma_{\rm SN}(t) = 0.0$.

The swept ISM mass by SN shocks $m_{\rm swept}$ depends on both the density
and metallicity of the ISM.
In the case of a higher density, since the amount of materials which
block SN blast wave is larger, the swept mass becomes smaller.
Further, the line cooling by metals is more efficient
in the ISM of a higher metallicity, as a result, the swept mass becomes smaller.
To consider these effects, we adopt the fitting formula derived by
Yamasawa et al. (2011)
\begin{equation}
m_{\rm swept} = 1535 n^{-0.202}_{\rm SN} \left[(Z/\mbox{Z}_{\odot}) + 0.039\right]^{-0.289}\;[\mbox{M}_{\odot}],
\end{equation}
where $n_{\rm SN}$ is the ISM density around SNe, we assume $n_{\rm SN}
= 1.0\;{\rm cm}^{-3}$ as a representative value.

\subsection{Metal accretion timescale}

Dust mass in galaxies increases due to not only supply from stars but
also accretion of refractory elements onto preexisting dust in clouds
(e.g., Liffman \& Clayton, 1989; Draine, 2009; Jones \& Nuth, 2011).
This accretion process is called ``dust mass growth''.
Here, we use a term ``clouds'', so that it stands the cool component in
the ISM.
In our study, we neglect volatile dusts.
Although, indeed, they exist in clouds, if clouds disappear or the
temperature goes up, such dusts evaporate. 
More precise treatment will be shown in future work.
The dust mass growth rate in clouds can be expressed as (e.g.,
Hirashita, 2000; Inoue, 2003, 2011)
\begin{equation}
\left(\frac{{\rm d}M_{\rm d}}{{\rm d}t}\right)_{\rm acc} = \eta N \pi
 \langle a^2 \rangle \alpha \rho^{\rm gas}_{Z} \langle v \rangle,
\end{equation}
where $\eta$ is the mass fraction of the clouds, $N$ is the number of
dust grains, $\langle a ^2 \rangle$ is the 2nd moment of a grain size $a$,
$\alpha$ is the mean sticking coefficient of metals, $\rho^{\rm gas}_{Z}$
is the mass density of gaseous metals that are not contained in dust,
and $\langle v \rangle$ is the mean velocity of metals in gas phase.
Since we assume a spherical dust grain for simplicity, 
we have
\begin{equation}
m_{\rm d} = \frac{4 \pi \langle a^3 \rangle \sigma}{3},
\end{equation}
so
\begin{equation}
N = \frac{M_{\rm d}}{m_{\rm d}} = \frac{3M_{\rm d}}{4 \pi \langle a^3 \rangle \sigma},
\end{equation}
where $m_{\rm d}$ is the mean mass of the grain, $\langle a^3 \rangle$
is the 3rd moment of a grain size, and $\sigma$ is the mass density of
solid matter within the grains.
Also,
\begin{equation}
\rho^{\rm gas}_{Z} = \rho^{\rm eff}_{\rm ISM}Z(1 - \delta),
\end{equation}
where $\rho^{\rm eff}_{\rm ISM}$ is the averaged mass density of the
interstellar clouds where the accretion process occurs and $\delta$ is
the dust abundance in the metal mass.
Thus, defining the accretion timescale as
\begin{equation}
\tau_{\rm acc} = \frac{4 \langle a^3 \rangle \sigma}{3 \langle a^2
 \rangle \alpha \rho^{\rm eff}_{\rm ISM} Z \langle v \rangle},
\end{equation}
we obtain the fourth term in right hand side of Eq.~(4).
Also, the mass density $\rho^{\rm eff}_{\rm ISM}$ is estimated in terms
of the hydrogen number density, $n_{\rm H}$, as $\rho^{\rm eff}_{\rm
ISM} = \mu m_{\rm H} n_{\rm H}$, where $\mu$ is the mean atomic weight
(we assume $\mu = 1.4$; i.e.\ the ratio of
the number of a hydrogen atom and a helium atom is $10:1$ in clouds),
and $m_{\rm H}$ is the mass of a hydrogen atom.

We assume that $\alpha = 1.0$ (i.e.\ if a molecule/atom collides a dust
grain, it sticks certainly) and $\sigma = 3\;\mbox{g\,cm}^{-3}$ (silicate).
Considering that the dust mass growth depends on the volume-to-surface
ratio of the grains, then we obtain 
\begin{eqnarray}%\label{eq:accretion2}
\nonumber&\tau_{\rm acc}& \thickapprox 2.0 \times 10^7\\
\nonumber &\times& \left(\frac{\bar{a}}{0.1{\rm \mu
m}}\right)\left(\frac{n_{\rm H}}{100\;\rm{cm}^{-3}}\right)^{-1}\left(\frac{T}{50\;{\rm
K}}\right)^{-\frac{1}{2}}\left(\frac{Z}{0.02}\right)^{-1}\hspace{-10pt}[{\rm yr}]\\
&=& \tau_{\rm acc,0}\;Z^{-1},
\end{eqnarray}
where $\tau_{\rm acc,0} = 4.0 \times 10^5$~yr, and we adopted $\bar{a} =
0.1\;\mu$m, $n_{\rm H} = 100\;{\rm cm}^{-3}$, and $T = 50\; {\rm
K}$~\footnote[1]{${}^{1}$This temperature corresponds to $\langle v \rangle
= 0.14$~km~${\rm s}^{-1}$. We assume $Am_{\rm H} \langle v \rangle^2 =
kT$ and adopt $A = 20$~($Am_{\rm H}$ is the mean mass of the colliding
atoms) (Spitzer, 1978).}.
The typical size of grains $\bar{a}$ is defined as $\langle a^{3}
\rangle/\langle a^{2} \rangle$ in Hirashita \& Kuo (2011).

As mentioned above, we conservatively adopt $\bar{a} = 0.1\;\mu$m as a
fiducial value (e.g., Inoue, 2011).
Small grains may be depleted by
coagulation in molecular clouds (Hirashita \& Yan, 2009), which strengthens
the importance of large grains.
The importance of large grains is further enhanced given that the grain
size distribution tends to be biased to large ($a \sim 0.1\;\mu$m) size
by the destruction within SN remnants (Nozawa et al., 2007).
Thus, we assume $\bar{a} \sim 0.1\;\mu$m to estimate the dust mass growth
timescale. 
Although we basically adopt $\bar{a} = 0.1\;\mu$m, we also
examine $\bar{a} = 0.01\;\mu$m for a quick
growth case later. Indeed, the MRN grain size distribution
(Mathis, Rumpl \& Nordsieck, 1977) has $\bar{a} = 0.01\;\mu$m (Hirashita
\& Kuo, 2011).
In reality, the grain size distribution in galaxies changes with time due to some
processes (e.g., SN shocks, accretion, and etc.).
As for the contribution of the evolution of the grain size distribution,
we have been preparing Asano et al. (2012).

In this paper, we adopt only $\eta = 0$ (no accretion growth) or $1$ in
order to avoid any fine-tuning.
In fact, the effect of a different choice of $\eta$ can be offset by a
different choice of $n_{\rm H}$ and $T$.
This allows us to merge uncertainties of $\eta$, $n_{\rm H}$, and $T$
into the value of $\tau_{\rm acc,0}$.
We set $\tau_{\rm acc,0} = 4.0 \times 10^{5}$~yr as a fiducial value
(e.g., Inoue, 2011).
Other choices of $\tau_{\rm acc,0}$ result in a different timing of the
growth activation.
This is explicitly expressed in Eq.~(27) later.

\section{What drives dust mass growth in the ISM?}
%\label{sec:result}

In this section, we investigate what determines the point where the grain
growth in the ISM becomes effective to the total dust mass in galaxies.

\subsection{Contribution of each physical process to the total dust mass in galaxies}
%\label{sec:contri}

{}To examine the point where the dust mass growth becomes efficient as a main source of
dust in galaxies, at first, we compare each process of dust production.
In Fig.~1, we show the contributions of stars (solid and
dotted lines), dust destruction (dashed line) and the dust mass growth (dot-dashed line) to the
total dust mass in a galaxy.
Solid, dotted, dashed and dot-dashed lines represents the contributes of the 1st,
2nd, 3rd and 4th terms in r.h.s. of Eq.~(4), respectively.
\begin{figure*}[t]
\centering\includegraphics[width=0.45\textwidth]{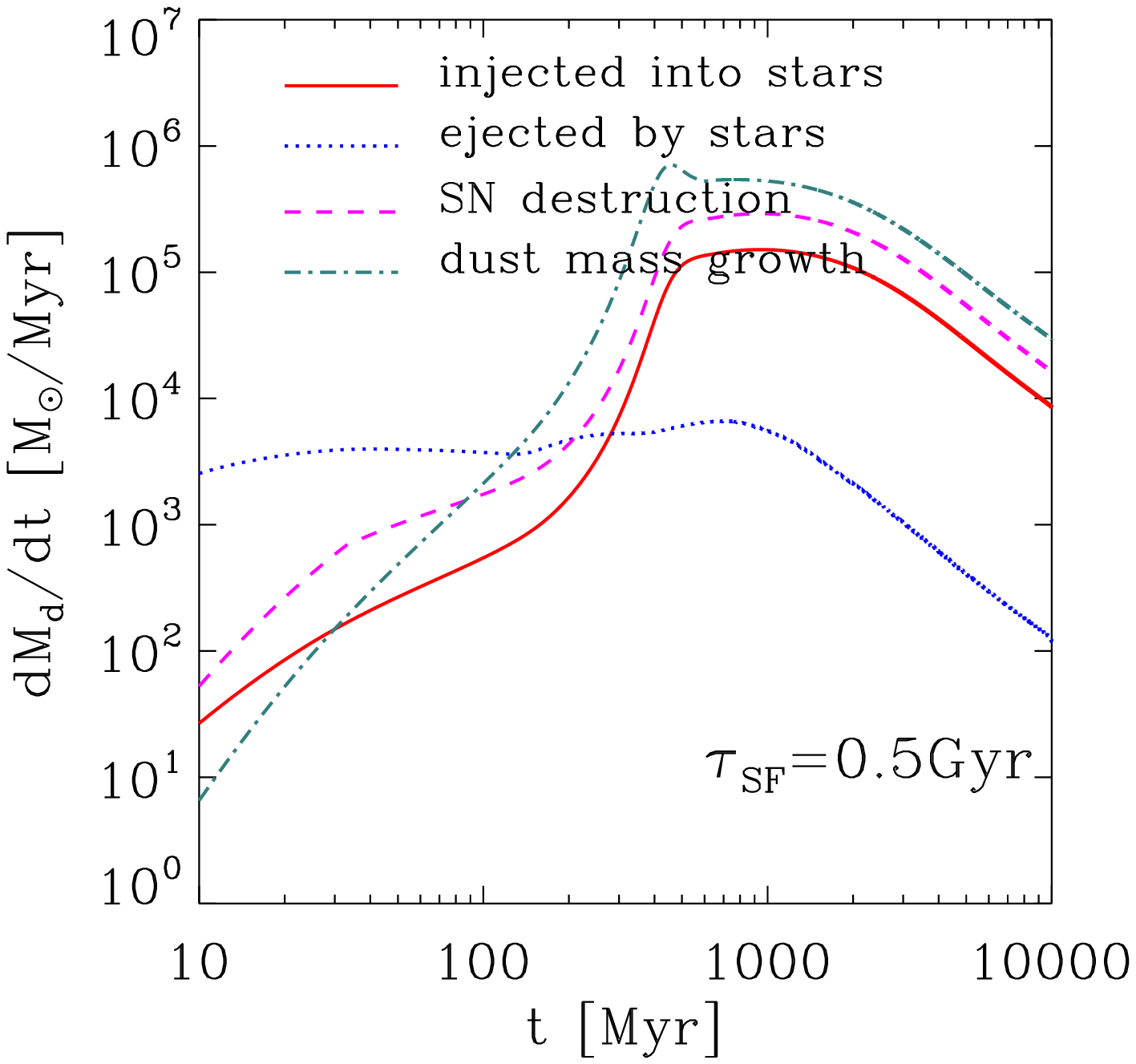}
\includegraphics[width=0.45\textwidth]{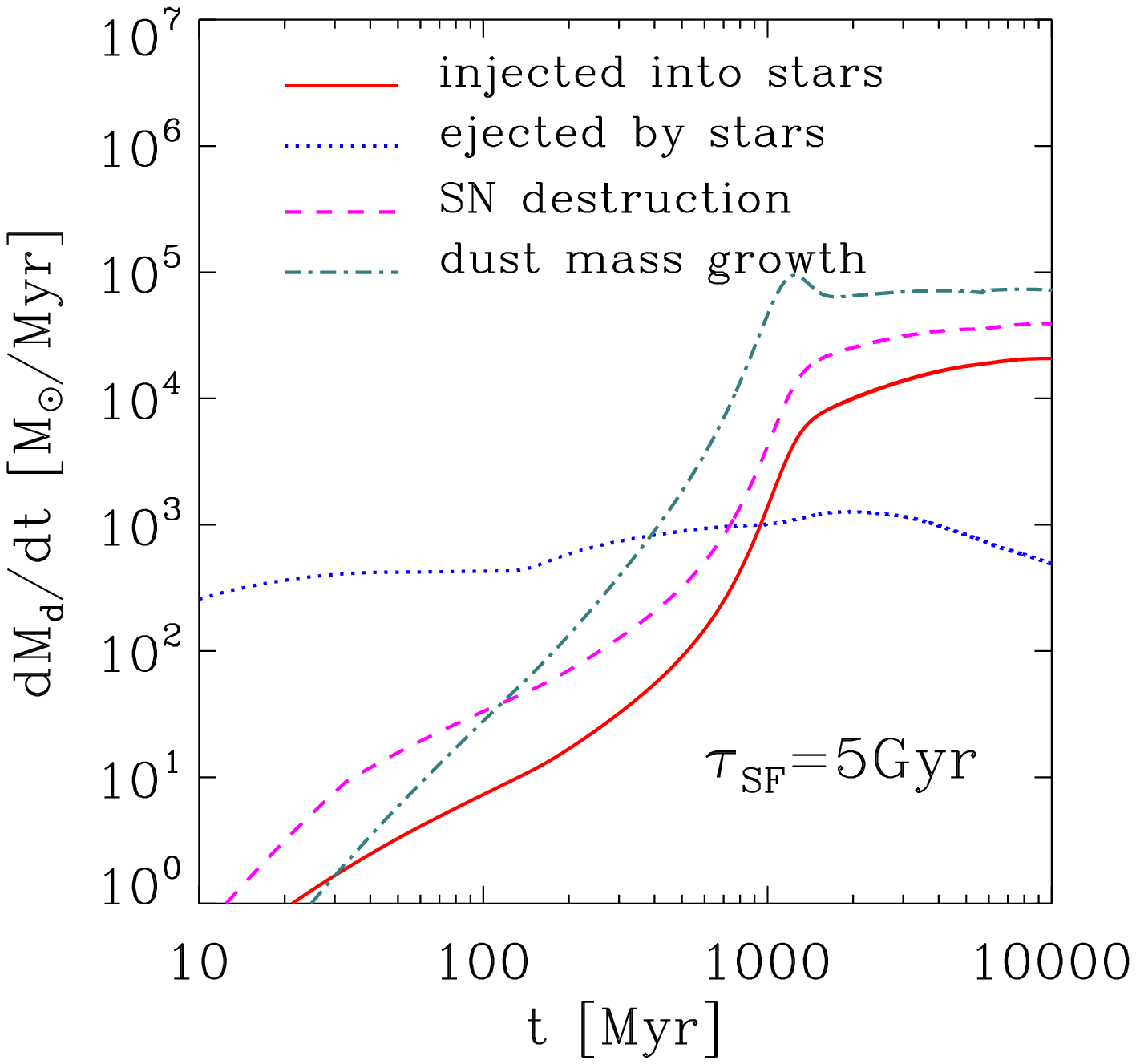}
\includegraphics[width=0.45\textwidth]{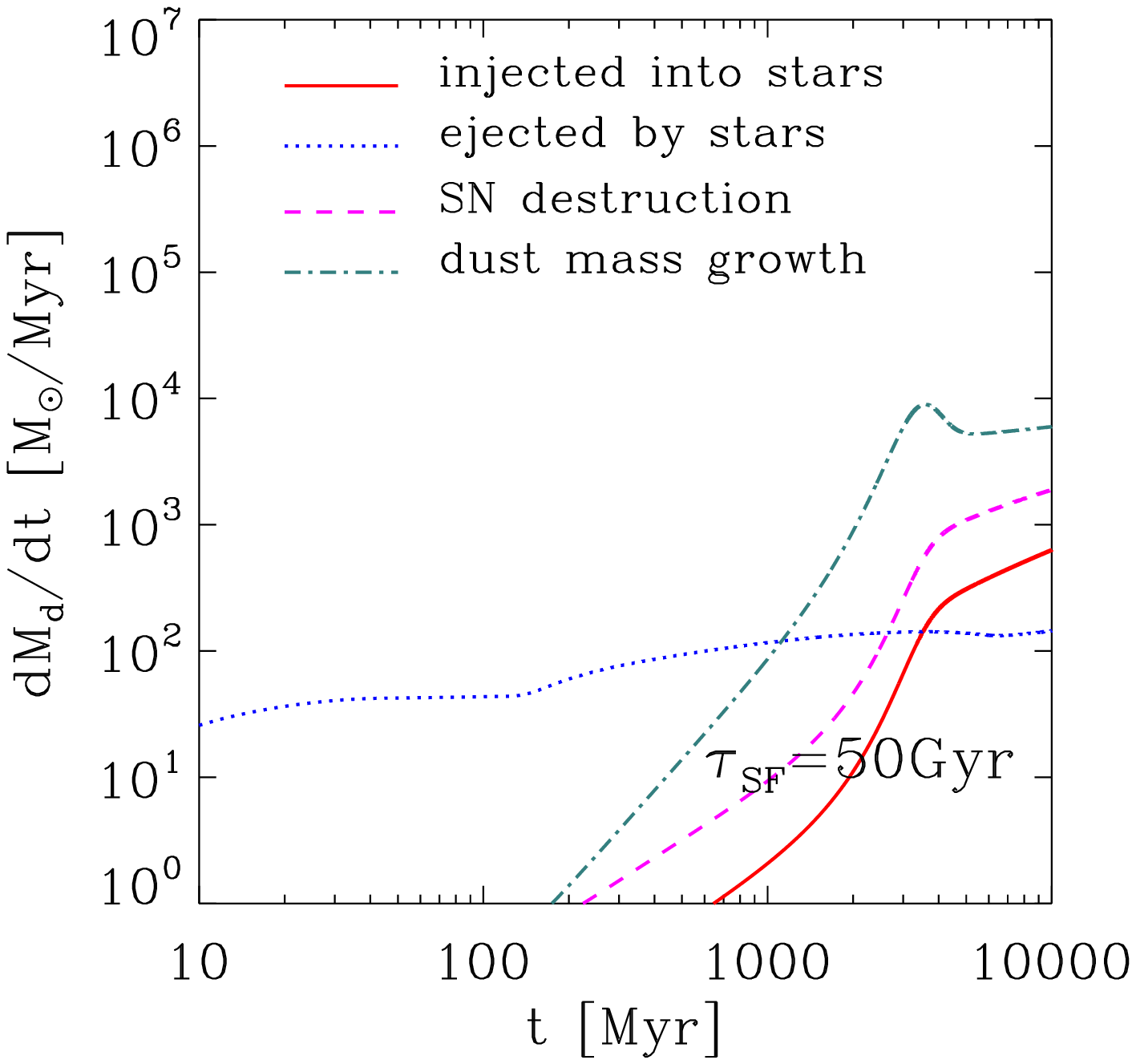}
\caption{Time evolution of each dust production/destruction rate [terms
 in right hand side of Eq.~(4)] with $M_{\rm tot} =
 10^{10}\;\mbox{M}_{\odot}$ and $\eta = 1.00$.
The star formation timescales are set to be $0.5$~Gyr (top-left panel),
 $5$~Gyr (top-right panel) and $50$~Gyr (bottom panel).
Solid, dotted, dashed and dot-dashed lines represent the rates of dust
 injection into stars, dust ejection by stars, dust destruction by SN
 shocks and dust mass growth in a galaxy, respectively.}
%\label{fig:diff}
\end{figure*}
The total baryon mass $M_{\rm tot}$ (the sum of the total mass of stars
and ISM in a galaxy) and $\eta$ are $10^{10}\;\mbox{M}_{\odot}$ and
$1.00$, respectively.
However, as mentioned in Sect.~2.1, since $M_{\rm tot}$ is
just a scale factor, $M_{\rm tot}$ just changes the values of the
contributions of these processes linearly. 

{}From these figures, we find that although the contribution of the ejected by stars is
the biggest in the early stages, as the time passes, the main contributor
of the dust production switches to the dust mass growth in the ISM at a point
(``switching point'').
For example, Liang \& Li (2009) pointed out that dust produced by SNe~II
is predominant to the dust budget in galaxies at high-$z$ Universe ($z >
5$) using the extinction curves of GRB host galaxies at high redshifts. 
Their results are in good agreement with this work.
Further, the process of the dust mass growth is expected to explain the dust amount
in the Milky Way or dusty QSOs at high redshifts
(e.g., Zhukovska, Gail \& Trieloff, 2008; Draine, 2009; Micha\l owski et
al., 2010a; Valiante et al., 2011).
So, what determines the switching point?
We will discuss the point in next subsection (this is the main topic in
this paper).

After the dust mass growth took place, the contribution of dust
destruction by SN shocks approaches that of the dust mass growth.
Thus, after the dust mass growth becomes efficient, dust amount in galaxies determines
the balance between the contribution of dust destruction and that of the
dust mass growth in the ISM (see also Inoue, 2011).

In addition, we also observe that the increase of the contribution of the dust mass growth
(dot-dashed line) has a peak, after that, the increase slows down.
In other words, the dust mass growth becomes ineffective.
We consider the reason of this.
In Fig.~2, we show the time evolution of $\delta$ ($= M_{Z}/M_{\rm d}$) with various star
formation timescales.
\begin{figure}[t]
\centering\includegraphics[width=0.45\textwidth]{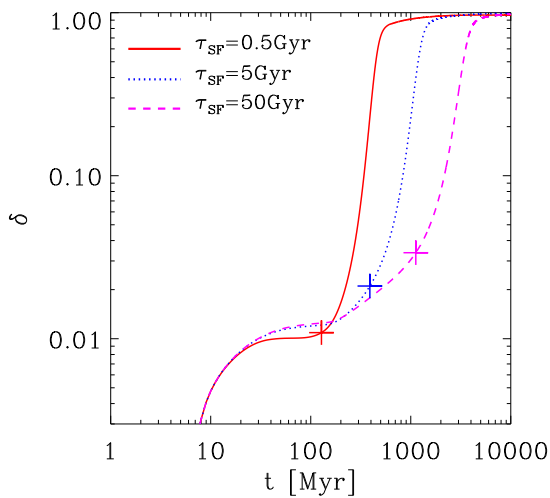}
\caption{Time evolution of the fraction of metals in dust, $\delta$ with
 $\eta = 1.00$. Solid, dotted and dashed lines represent $\tau_{\rm SF}
 = 0.5, 5, 50$~Gyr, respectively.
Cross symbols mark the switching point for each $\tau_{\rm SF}$.}
%\label{fig:delta}
\end{figure}
We find that after the value of $\delta$ increases rapidly, the value
saturates.
Hence, after the dust mass growth becomes effective, most of metals form
dust.
Thus, the reason why the dust mass growth becomes ineffective is the
depletion of metals.

As shown in Fig.~2, the values of $\delta$ for all $\tau_{\rm SF}$s
converge to $\sim 1$.
In contrast, the value for the Milky Way is about $0.5$.
However, since it can be adjusted by adopting different $\eta$, we do
not try to fine-tune the convergence value of $\delta$ in this study.
Inoue (2011) showed that the convergence value of $\delta$ is
determined by the balance between the contribution of dust destruction
by SN shocks and that of the dust mass growth (for details, the product
of $\tau_{\rm acc,0}$ and $\epsilon m_{\rm swept}$).

\subsection{Critical metallicity for dust mass growth}
%\label{sec:crimetal}

In this subsection, we introduce the main topic of this paper, the critical metallicity.
This is a metallicity at the switching point (see Sect.~3.1).
In our model, the sources of dust are stars (AGB stars
and SNe~II) and the dust mass growth in the ISM. 

In order to find the critical metallicity, we compare the second term 
with fourth term in right hand side of Eq.~(4).
First we consider the second term [Eq.~(9)]. 
{}From Eq.~(9), if $D$ is defined as 
\begin{equation}
\int^{100\;{\rm M}_{\odot}}_{m_{\rm cut}(t)} m_{\rm d}(m,Z(t - \tau_m))\phi(m){\rm d}m \equiv D, 
\end{equation}
then Eq.~(9), with Eq.~(5) can be approximated as
\begin{equation}
Y_{\rm d} \simeq D\frac{M_{\rm ISM}}{\tau_{\rm SF}}.
%\label{eq:app_star}
\end{equation}
This equation is exact if SFR is constant.

Next, we consider the fourth term in right hand side of
Eq.~(4). 
Since $\tau_{\rm acc} = \tau_{\rm acc,0}\;Z^{-1}$ and $M_{\rm d} = \delta
M_Z = \delta Z M_{\rm ISM}$, the dust mass growth term is  
\begin{equation}
\eta\frac{M_{\rm d}(1 - \delta)}{\tau_{\rm acc}} = \frac{\eta \delta(1 -
 \delta)Z^2 M_{\rm ISM}}{\tau_{\rm acc,0}}.
%\label{eq:app_growth}
\end{equation}

{}From Eq.~(22) and (23), the
metallicity at which the increasing rate of dust mass by the dust mass growth exceeds the dust
production rate by stars is, then, presented as follows:
\begin{equation}
Z = \left[ \frac{D}{\eta \delta(1 - \delta)}\right]^{\frac{1}{2}}
 \left(\frac{\tau_{\rm acc,0}}{\tau_{\rm SF}}\right)^{\frac{1}{2}}.
%\label{eq:crimetal}
\end{equation}
Thus, if the metallicity of a galaxy is larger than above metallicity,
 we should consider the effect of dust mass growth in the galaxy.
Here, we refer to the metallicity as \textit{the critical metallicity}
 $Z_{\rm cr}$, which is the metallicity at the switching point.
To obtain the value of $Z_{\rm cr}$, hereafter, we adopt $\delta = 0.02$ and $D = 5 \times 10^{-4}$.
 As for the value of $\delta$, from Fig.~2, the
 value of $\delta$ ranges $0.01$--$0.04$ at the switching point for each $\tau_{\rm SF}$.
 Further, although $\delta$ is dependent on time, before the
 dust mass growth becomes effective to the total dust 
 mass, the $\delta$ is determined only by the contribution of stars
 (see Appendix~A).
 This contribution of stars ranges $0.01$--$0.04$ in our calculation
 (Fig.~8 in Appendix~A). 
Thus, we adopt $\delta = 0.02$ as a representative value. 
Also, since we found from numerical calculation that the range of $D$ is
$10^{-4}$ to $10^{-3}$, we adopt $D = 5 \times 10^{-4}$ as a representative
value.

For the reader's convenience, we compare our discussion with a similar
work by Inoue (2011).
Inoue (2011) defined a critical metallicity to compare the
contribution of the dust destruction by SN shocks with that of the dust mass
growth.
Thus, the critical metallicity in Inoue (2011) is a metallicity which
the contribution of the dust mass growth exceeds that of the dust
destruction.
In contrast, our critical metallicity is a metallicity which the dust
mass growth becomes the main source of the increase of dust (the contribution
of dust mass ejected by stars is the main source of dust at early
stage of galaxy evolution).
Readers who are interested in both works should keep this difference in mind.

In Fig.~3, we show the relation between the
metallicity and dust-to-gas mass ratio (left panel is normalized by the
critical metallicity, while right panel is not normalized) for $\tau_{\rm SF} = 0.5, 5, 50$~Gyr.
\begin{figure*}[t]
\centering\includegraphics[width=0.45\textwidth]{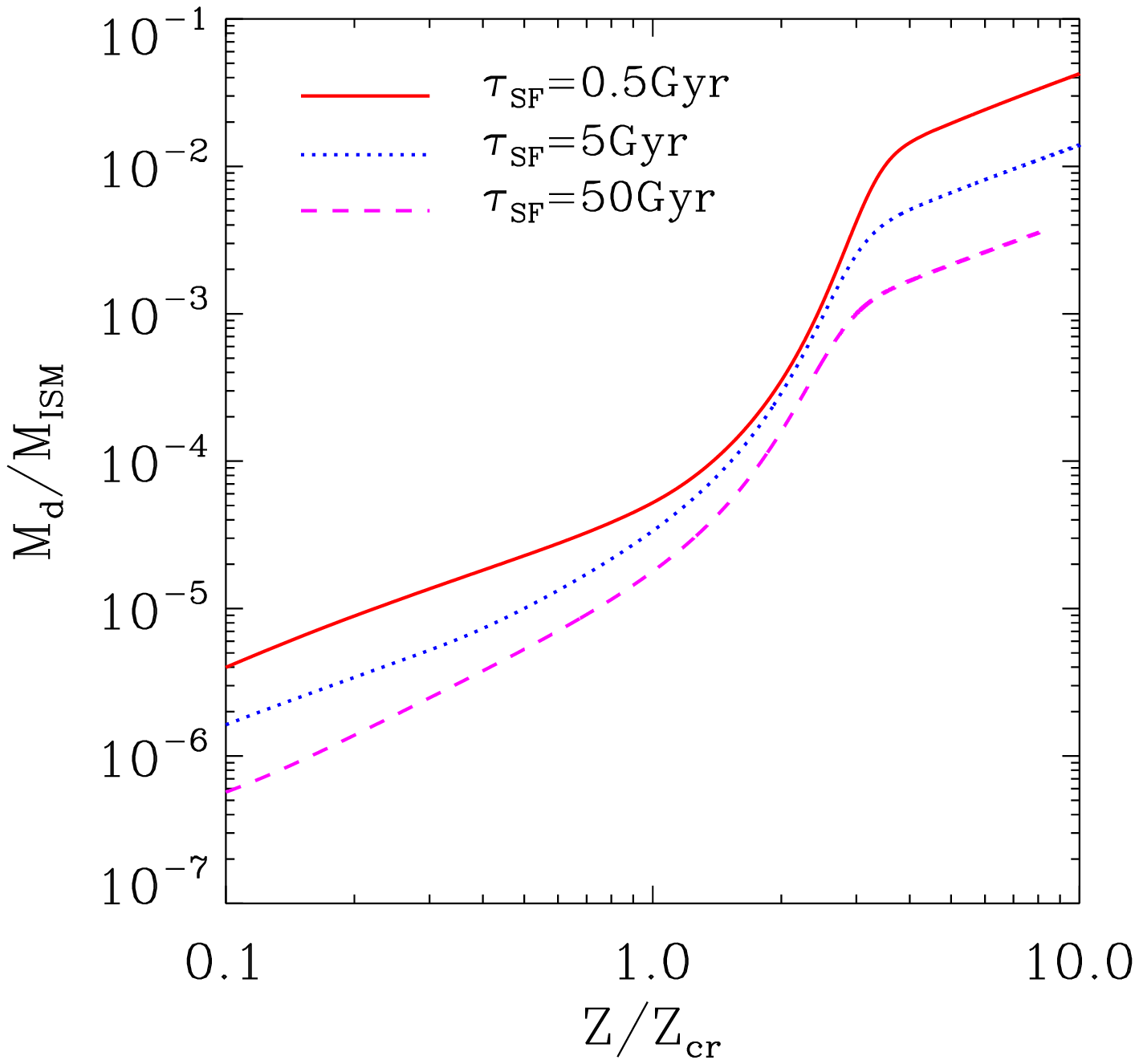}
\includegraphics[width=0.45\textwidth]{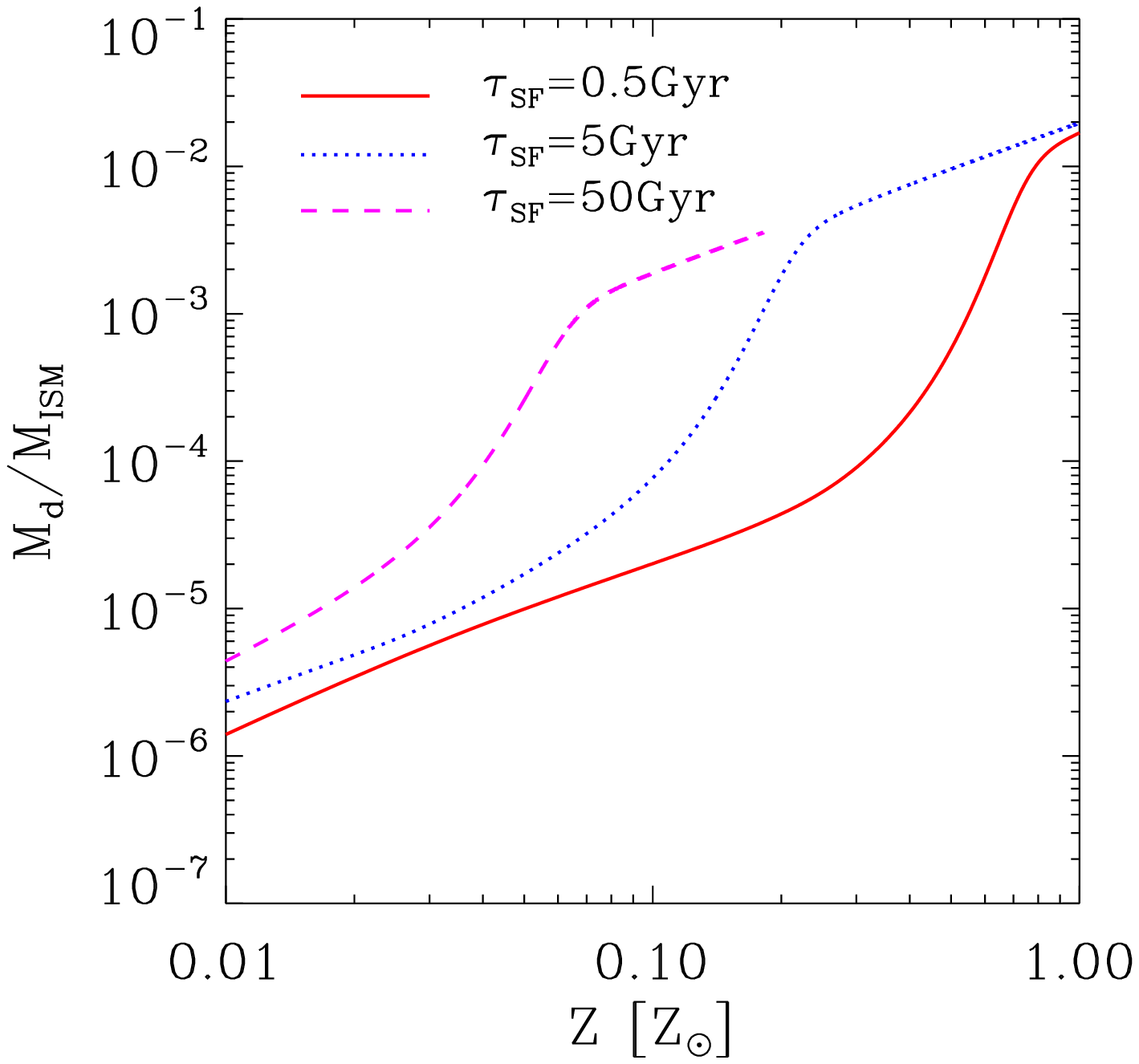}
\caption{The metallicity - dust-to-gas mass ratio with $\eta = 1.00$
 and $M_{\rm tot} = 10^{10}\;{\rm M}_{\odot}$. 
Left panel: normalized by the critical metallicity.
Right panel: not normalized.
Solid, dotted and dashed lines represent $\tau_{\rm SF} = 0.5, 5, 50$~Gyr, respectively. 
Dashed and solid horizontal lines represent $\beta$ = 1.0 and 10.0,
 respectively.}
%\label{fig:crimetal}
\end{figure*}
In Fig.~3, we find different evolutionary tracks
depending on $\tau_{\rm SF}$ in right-hand panel, whereas these tracks
with metallicity normalized to the critical one are well overlaid on each other in the left-hand panel.
This clearly shows that the critical metallicity is truly essential: The
dust mass growth becomes efficient not when the galactic age reaches a
certain value but when metallicity exceeds the critical value.
In the left panel, the dust-to-gas mass ratio increases rapidly after
metallicity exceeds the point of the critical metallicity.
Thus, dust produced by the dust mass growth becomes efficient to the dust
mass of a galaxy if $Z > Z_{\rm cr}$ in the galaxy.

In Fig.~4, we show the critical metallicity as a
function of $\tau_{\rm SF}$.
As shown in the figure, the critical metallicity becomes larger if the
star formation timescale is shorter.
As mentioned in Sect.~1 and 3.1, the dust mass growth is
expected to be the dominant source of dust in various galaxies (e.g.,
the Milky Way: Draine (2009) and dusty QSOs at high
redshifts\footnote[2]{${}^{2}$It is considered that QSOs in high-$z$ Universe have high star formation rate
and even and larger subsolar metallicities (e.g., Matsuoka et al., 2009;
Juarez et al., 2009).}:
Valiante et al. (2011), among others),
in spite of different star formation timescales and ages of the
galaxies.
In terms of the critical metallicity, we can explain the reason of it in
a coherent manner; metallicity in the galaxies just exceeds the critical one.
\begin{figure}[t]
\centering\includegraphics[width=0.45\textwidth]{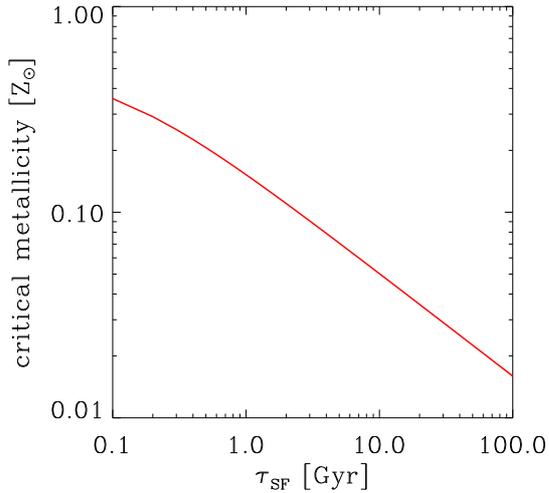}
\caption{Critical metallicity as a function of $\tau_{\rm SF}$ with $\eta = 1.00$.}
%\label{fig:crimetal2}
\end{figure}

\begin{figure}[t]
\centering\includegraphics[width=0.45\textwidth]{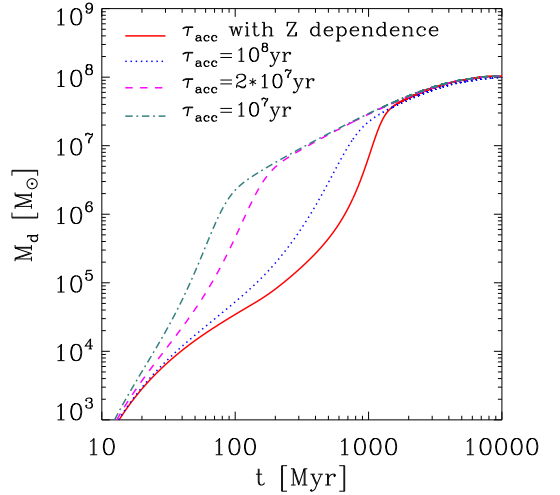}
\caption{Time evolution of the total dust mass with (solid line)
 and without (other lines) metallicity dependence for the
 dust mass growth timescale. Here we
 adopt $\tau_{\rm SF} = 5$~Gyr, $M_{\rm tot} = 10^{10}\;{\rm
 M}_{\odot}$ and $\eta = 1.00$. The three values of $\tau_{\rm acc}$
 represent $Z = 0.2\;Z_{\odot}$ (dotted line), $Z = 1.0\;Z_{\odot}$
 (dashed line) and $Z = 2.0\;Z_{\odot}$ (dot-dashed line), respectively.}
%\label{fig:crimetaleff}
\end{figure}
Here we emphasize the importance of the metallicity dependence in the
accretion growth timescale, $\tau_{\rm acc}$.
Pipino et al. (2011) argued that the dust mass growth was important to explain the
observed huge mass of dust in high-$z$ QSOs.
However, they did not seem to consider the metallicity dependence in
$\tau_{\rm acc}$\footnote[3]{${}^{3}$Indeed, their equations (4) and (5) have the
metallicity dependence. However, their adopted timescales in table 1 seem
to omit the dependence finally.}.
Fig.~5 shows the effect of the metallicity
dependence on the dust mass evolution.
When we consider the dependence properly, the timing where the dust mass growth
becomes effective is delayed until the metallicity exceeds the critical value as discussed
above.
On the other hand, if we omit the dependence and adopt a constant value
for $\tau_{\rm acc}$, the timing of the growth becomes effective on the
total dust mass is determined by just the
adopted $\tau_{\rm acc}$.

In the above discussion, we have focused on the critical metallicity.
One may be, however, interested in its relation to the time, $t_{\rm cr}$,
which is a galactic age when the metallicity in a galaxy reaches the critical metallicity.
Here, in order to understand the importance of the dust mass growth in
various galaxies with various star formation timescale more clearly,
we demonstrate the relation between the critical metallicity
$Z_{\rm cr}$ and the time $t_{\rm cr}$.
However, we stress that the metallicity is more fundamental because
{\it $t_{\rm cr}$ is determined by the critical metallicity.}

We derive the relation with using Eq.~(2) and
(3).
If $A$ is defined as
\begin{equation}
\int^{100\;{\rm M}_{\odot}}_{m_{\rm cut}(t)} m_Z(m,Z(t - \tau_m))
 \phi(m) {\rm d}m \equiv A,
\end{equation}
then, the relation is expressed as
\begin{equation}
t_{\rm cr} = \frac{\tau_{\rm SF}}{A}Z_{\rm cr},
%\label{eq:critime}
\end{equation}
where A is a constant, and is about 0.018 in our calculation.
If inflow process occurs, since dust-to-gas mass ratio and metallicity
become smaller than the case without inflow, 
it is sufficient to consider the larger $\tau_{\rm SF}$.
Figure~6 shows $t_{\rm cr}$ as a function of
$\tau_{\rm SF}$.
From this figure, we find that $t_{\rm cr}$ becomes shorter if
$\tau_{\rm SF}$ is shorter, while in the case of $Z_{\rm cr}$, the trend
is opposite (Fig.~4).
This is explained by following reason.
If the $\tau_{\rm SF}$ is short, the fast progress of the star formation
makes the metallicity become large at an early stage of galaxy
evolution.
Hence, although the critical metallicity is large in the case with short
$\tau_{\rm SF}$, $t_{\rm cr}$ is short due to the fast evolution of
metallicity.
From the relation between $z_{\rm cr}$ and $t_{\rm cr}$, we can
understand that the dust mass growth is the main dust production of the
Milky Way (evolved galaxy) and dusty QSOs (in high-$z$ Universe).

\begin{figure}[t]
\centering\includegraphics[width=0.45\textwidth]{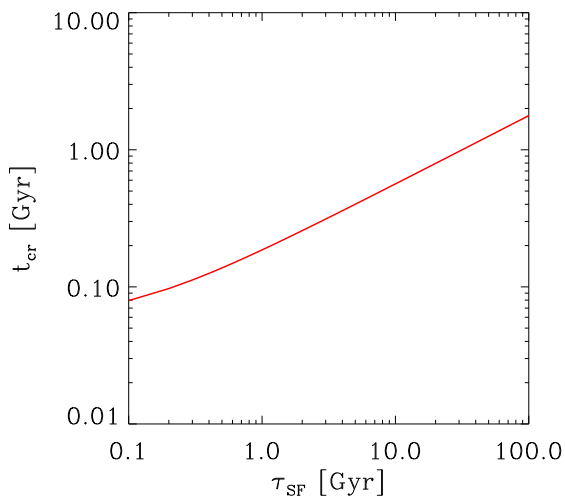}
\caption{$t_{\rm cr}$ as a function of $\tau_{\rm SF}$ with $\eta = 1.00$.}
%\label{fig:criticaltime}
\end{figure}

Next, we discuss the effect of the grain size distribution.
Recently, Hirashita \& Kuo (2011) showed that the dust mass growth in the ISM
depends on the grain size distribution of a galaxy.
{}To examine the effect of the grain size on the dust mass growth, we consider the case with
$\bar{a} = 0.01\;\mu$m (the case with $\bar{a} = 0.1\;\mu$m is the
fiducial case in this paper).
Figure 7 shows the relation between metallicity and
dust-to-gas mass ratio for $\bar{a} = 0.1\;\mu$m (solid line) and
$0.01\;\mu$m (dotted line) with $\tau_{\rm SF} = 5$~Gyr and $\eta = 1.00$.
{}From Eq.~(20) and (24), the critical
metallicity as a function of $\bar{a}$ is expressed as
\begin{equation}
%\label{eq:crichange}
Z_{\rm cr}(\bar{a}) =(\bar{a}/0.1\;\mu\rm{m})^{1/2} Z_{\rm cr}(0.1\;\mu\rm{m}).
\end{equation}
{}From Fig.~7, we find that the
evolutionary tracks with different $\bar{a}$ show the almost
same behavior if we introduce the critical metallicity for each value of
$\bar{a}$.
Thus, although Hirashita \& Kuo (2011) showed that the critical metallicity is
sensitive to the grain size distribution, the mechanism that the critical metallicity
determines the timing on which the dust mass growth becomes dominant to the growth
of the total dust mass in a galaxy does not change.
As for the dust evolution considered the evolution of the grain size
distribution (including the effects of the stellar dust, SN destruction
and accretion) in a galaxy,
This issue will be extensively discussed in our next work 
(Asano et al., 2012, in preparation).

\begin{figure}[t]
\centering\includegraphics[width=0.45\textwidth]{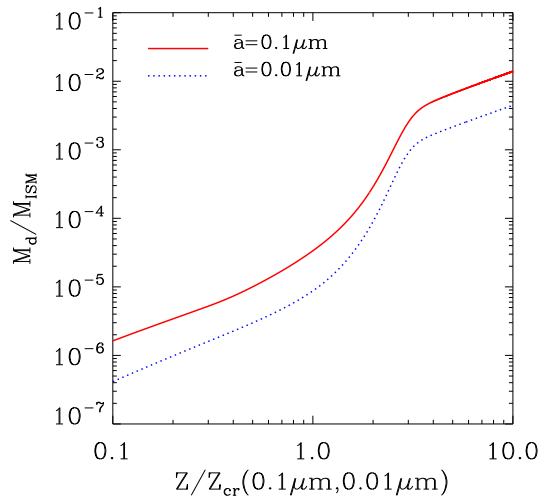}
\caption{The relation between metallicity and
dust-to-gas mass ratio for $\bar{a} = 0.1\;\mu$m (solid line) and
$0.01\;\mu$m (dotted line) with $\tau_{\rm SF} = 5$~Gyr and $\eta = 1.00$.
{}From Eq.~27, $Z_{\rm cr}(0.01\;\mu{\rm m}) \sim
 0.32\;Z_{\rm cr}(0.1\;\mu{\rm m})$.}
%\label{fig:crimetalsize}
\end{figure}

\section{Conclusions}
%\label{sec:conclusion}

In this work, we constructed a galaxy evolution model taking
into account the metallicity and age dependence on the various dust sources (AGB stars, SNe~II and
growth in the ISM) to investigate what is the main driver of the grain
growth which is expected to be the dominant source of dust in various
galaxies with various star formation timescales.

We have found that the timing that the dust mass growth in the ISM
becomes effective is determined by metallicity.
If metallicity in a galaxy exceeds a certain critical value, \textit{critical metallicity},
the dust mass growth becomes active and the dust mass rapidly increases
until metals are depleted from the ISM.
This critical metallicity is larger for a shorter star formation
timescale.
The dust mass growth is thought to be the dominant source of dust in evolved
galaxies like the Milky Way and young but dusty and massive QSOs at high redshifts. 
The importance of the dust mass growth in such a diversity of galaxies can
be explained clearly in terms of the critical metallicity;
the dust mass growth in the ISM is regulated by metallicity, and
we stress that the critical metallicity works as an indicator to judge whether the grain
growth in the ISM is dominant source of dust in a galaxy,
especially because of the strong and nonlinear dependence on the 
metallicity.

%\subsubsection{Subsubsection}\strut

\acknowledgments{
We thank the anonymous referees for their helpful comments which improved the presentation and content
of this paper.
We are grateful to Takashi Kozasa, Takaya Nozawa,
Daisuke Yamasawa, Asao Habe, and Takako T.\ Ishii 
for fruitful discussions.
RSA has been supported from the Grant-in-Aid
for JSPS Research under Grant No. 23-5514.
RSA and TTT have been also partially supported from the Grand-in-Aid for the Global 
COE Program ``Quest for Fundamental Principles in the Universe: from 
Particles to the Solar System and the Cosmos'' from the Ministry of
 Education, Culture, Sports, Science and Technology (MEXT) of Japan.
TTT and AKI have been supported by Program for Improvement of Research 
Environment for Young Researchers from Special Coordination Funds for 
Promoting Science and Technology, and the Grant-in-Aid for the Scientific 
Research Fund (TTT: 20740105, 23340046, AKI: 19740108) commissioned by the MEXT.
HH is supported by NSC grant 99-2112-M-001-006-MY3.}

\appendix{What dominates $\delta$ before the dust mass growth becomes effective?}
%\label{secapp:delta}

Figure 8 shows the time evolution of the ratio of dust mass produced by stars to
metal mass ejected by stars, $\delta_{\rm star}$ (thick lines) and
$\delta$ (thin lines) with $\tau_{\rm SF} = 0.5, 5, 50$~Gyr.
From Eq.~(7)--(9), the ratio $\delta_{\rm star}$ is expressed as 
\begin{equation}
\delta_{\rm star}(t) = \frac{Y_{\rm d}(t)}{R_{Z}(t) + Y_{Z}(t)}.
\end{equation}
From Fig.~8, we observe that the evolutionary tracks of $\delta_{\rm star}$ are the
almost same tracks before the dust mass growth becomes effective to the
total dust mass in galaxies.
Hence, $\delta$ before the dust mass growth becomes effective is
determined by $\delta_{\rm star}$.

\begin{figure}[t]
\centering\includegraphics[width=0.45\textwidth]{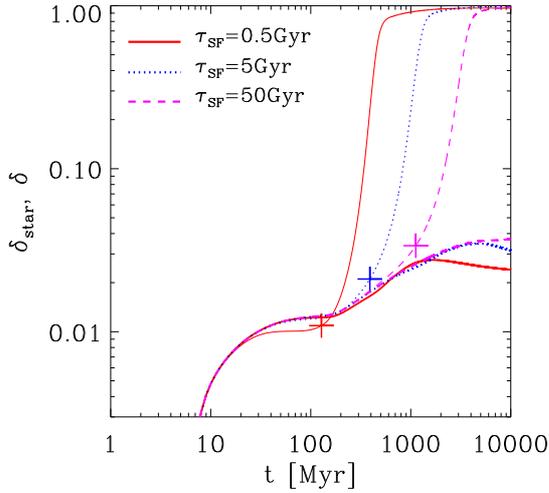}
\caption{Time evolution of the ratio of dust mass produced by stars to
 metal mass ejected by stars. Solid, dotted and dashed lines represent
 $\tau_{\rm SF} = 0.5, 5, 50$~Gyr, respectively.
For comparison, the results of Fig.~2 with
 thin lines are overlaid on the panel.
Cross symbols mark the switching point for each $\tau_{\rm SF}$.}
%\label{fig:stardelta}
\end{figure}

\email{R.S.Asano (e-mail: asano.ryosuke@g.mbox.nagoya-u.ac.jp)}
%\email{X.Xxxx (e-mail: xxxx@xxxx.xxxx.xxxx)}
\label{finalpage}
\lastpagesettings
\end{document}